\documentclass[conference, twocolumn]{IEEEtran}

\usepackage[utf8]{inputenc}
\usepackage{microtype}

\usepackage{amssymb}
\usepackage{graphicx}
\usepackage[hyphens]{url}
\usepackage[inline]{enumitem}
\usepackage{balance}
\usepackage{cite}

\usepackage{eurosym}
\usepackage{enumitem}
\usepackage{booktabs}
\usepackage[table,xcdraw]{xcolor}

\usepackage[flushleft]{threeparttable}
\usepackage{balance}
\usepackage{dcolumn}
\usepackage{longtable,tabu}
\usepackage{threeparttablex}
\usepackage{adjustbox}
\usepackage{tabularx}
\usepackage[underline=true]{pgf-umlsd}
\usetikzlibrary{calc}
\usetikzlibrary{arrows.meta}

\usepackage{arydshln}
\usepackage{multirow}

\usepackage{pifont}
\usepackage[caption=false]{subfig}

\usepackage[inline]{trackchanges}
\addeditor{Raja}
\addeditor{Kostas}
\addeditor{James}
\usepackage[mathlines,switch]{lineno}
\title{\LARGE \bf
An evaluation of the security of the Bitcoin Peer-to- Peer Network
}

\usepackage{graphicx}
\usepackage{makecell}
\usepackage{color}

\author{\IEEEauthorblockN{James Tapsell, Raja Naeem Akram, and Konstantinos Markantonakis}
\IEEEauthorblockA{ISG-SCC, Royal Holloway, University of London, Egham, United Kingdom\\
Email: \{James.Tapsell.2015\}@live.rhul.ac.uk, \{r.n.akram, k.markantonakis\}@rhul.ac.uk}}

\begin{document}
% START Triple line                                        -
% END Triple line                                          -
% START Jrtapsell Custom                                   -
\maketitle
\thispagestyle{empty}
\pagestyle{empty}
\graphicspath{ {images/} }
\newcommand{\comment}[1]{\textcolor{red}{\textbf{#1}}}
\newcommand{\jnumber}[1]{\textit{\#}\textbf{#1}}
\newcommand{\jfig}[2]{
	\begin{figure}[ht]
		\caption{#1}
		\centering
        \includegraphics[width=\columnwidth]{#2}
	\end{figure}
}
% END Jrtapsell Custom                                     -
% START Triple line                                        -
% END Triple line                                          -

%%%%%%%%%%%%%%%%%%%%%%%%%%%%%%%%%%%%%%%%%%%%%%%%%%%%%%%%%%%%%%%%%%%%%%%%%%%%%%%%
\begin{abstract}

Bitcoin is a decentralised digital currency that relies on cryptography rather than trusted third parties such as central banks for its security\cite{bitcoin_original}. Underpinning the operation of the currency is a peer-to-peer (P2P) network that facilitates the execution of transactions by end users, as well as the transaction confirmation process known as bitcoin mining. The security of this P2P network is vital for the currency to function and subversion of the underlying network can lead to attacks on bitcoin users including theft of bitcoins, manipulation of the mining process and denial of service (DoS). As part of this paper the network protocol and bitcoin core software are analysed, with three bitcoin message exchanges (the connection handshake, GETHEADERS/HEADERS and MEMPOOL/INV) found to be potentially vulnerable to spoofing and use in distributed denial of service (DDoS) attacks. Possible solutions to the identified weaknesses and vulnerabilities are evaluated, such as the introduction of random nonces into network messages exchanges.
\end{abstract}

%%%%%%%%%%%%%%%%%%%%%%%%%%%%%%%%%%%%%%%%%%%%%%%%%%%%%%%%%%%%%%%%%%%%%%%%%%%%%%%%

\section{Introduction}
Bitcoin operates as a currency through a peer-to-peer (P2P) network of nodes that execute, communicate and confirm transactions. Trust in its security is maintained by both the cryptographic elements of the system and the correct functioning of the P2P network.

\subsection{Contributions}
The key contributions of this paper are:
\begin{enumerate}
\item Analysis of two message exchanges in the bitcoin P2P network protocol (GETHEADERS and MEMPOOL) and their potential for spoofing and abuse in denial of service attacks.
\item Analysis of the security of hardcoded DNS seed addresses that allow nodes to first connect to the network.
\item Proposal of potential improvements to the security of the bitcoin P2P network protocol.
\end{enumerate}

\section{Bitcoin peer-to-peer network}
\subsection{Objectives \& challenges}
The overall goal of bitcoin, as put forward by Nakamoto, was to enable two entities to execute a transaction without relying on a trusted third party\cite{bitcoin_original}. The decentralised transmission of data (i.e. blocks and transactions) in bitcoin is carried out via a distributed peer-to-peer (P2P) network\cite{bitcoin_network}

To summarise the objectives of the bitcoin network:

\begin{table}[ht]
\caption{The objectives of the bitcoin network}
\label{table_example}
\begin{center}
\begin{tabular}{|c|c|}\hline
Objective & Achieved by\\\hline
\makecell{Data transmission must\\be decentralised\cite{bitcoin_network}} & \makecell{A distributed peer-to-peer network}\\\hline
\makecell{Data storage must\\be decentralised\cite{bitcoin_network}} & \makecell{A full copy of the blockchain\\is stored and maintained by all nodes}\\\hline
\makecell{All blocks must\\be accessible to all users\cite{bitcoin_original}} & \makecell{Blocks are broadcast to all nodes}\\\hline
\makecell{All transactions must\\be accessible to all users\cite{bitcoin_original}} & \makecell{Transactions are broadcast to all nodes}\\\hline
\end{tabular}
\end{center}
\end{table}

A key challenge with achieving these objectives is that nodes do not trust each other. Nodes must have the capability to verify information themselves without relying on a trusted third party.

\subsection{Network architecture}

The distributed P2P network is created in a dynamic way by users of bitcoin currency\cite{bitcoin_network}. 

Network nodes are homogeneous, with no specialised coordinating nodes and each node keeps a complete copy of the blockchain. This allows nodes to verify the validity of transactions and blocks independently without trusting each other\cite{bitcoin_propagation}.

Bitcoin nodes are identified by their IP address and operate over TCP (Transmission Control Protocol)\cite{bitcoin_deanonymisation}, which provides a reliable channel for bitcoin messages to be transmitted between nodes (i.e. guaranteed in-order delivery and recovery from transmission errors). 

However, there are no further security services beyond those provided by TCP, so bitcoin messages do not have cryptographic entity authentication or integrity protection.

\subsubsection*{Limitations of TCP}

TCP is a well-established and widely studied protocol and it has been known since 1989\cite{tcp_problems} that without any other cryptographic protection, it is trivial for an on path attacker (i.e. one that is situated on the communication path) to eavesdrop, modify, replay and fabricate TCP network packets.

An off path attacker (i.e. one that is not situated on the communication path) in a TCP/IP network can manipulate routing information to position themselves on the communications path and become an on path attacker (for example by manipulating the Routing Information Protocol\cite{tcp_problems} or Border Gateway Protocol\cite{bgp_problems}).

Furthermore off path attackers are still able to fabricate TCP packets by exploiting vulnerabilities in the use and selection of TCP sequence numbers\cite{tcp_weakness}, through various side channel attacks\cite{tcp_off}\cite{tcp_sequence}\cite{tcp_sequence2} and vulnerabilities in network implementations\cite{idle_scanning}\cite{tcp_offpath}.

In summary this means that, without any additional security mechanisms, an on path attacker can eavesdrop, modify, replay and fabricate TCP messages. Also if the TCP sequence number can be compromised\cite{tcp_problems}\cite{tcp_weakness}\cite{tcp_sequence}\cite{tcp_offpath}\cite{tcp_sequence}\cite{challenge_response}, an off path attacker can fabricate messages and impersonate another network node. 

As well as broadcasting transactions and blocks within TCP packets, bitcoin nodes also send TCP packets containing command messages. These command messages are used to establish and maintain connections between nodes and transfer data.

These messages are open to manipulation by network attackers as they are simply passed over TCP and do not provide any additional cryptographic protection.

\subsubsection{Initial connection}

To join the network for the first time a node discovers other nodes through DNS queries. The DNS names of several seed servers are hardcoded into the bitcoin core client software in the chainparams.cpp file\cite{bitcoin_chainparams}.

This provides a mechanism for nodes to connect to at least one peer, which will then provide them with further active peers to connect to. In this way the hardcoded DNS seed addresses act as the trusted, authoritative source for initial peers. After that, as the node interacts on the network it builds up a local database of active peers. 

The hardcoded DNS seed addresses are owned and managed by volunteers and have been chosen by the Bitcoin developers. Each query returns multiple IP addresses which correspond to bitcoin nodes that have high uptime. How these nodes are chosen and who manages them is not documented and raises a number of security concerns.

Were these DNS seeds to be compromised, an attacker could for example supply the addresses of their own malicious nodes to new nodes joining the network. Similar to an eclipse attack\cite{bitcoin_eclipse}, this would allow an attacker to supply the victim nodes with other malicious nodes to connect to and monopolise their network connections. In this way the DNS seeds would provide an additional vector for conducting an eclipse attack.

For example, an attacker could ensure that the addresses of malicious nodes are returned from the DNS seeds by:

\begin{itemize}
\item Exploiting DNS protocol weaknesses such DNS cache poisoning\cite{dns_security}\cite{dns_weakness} to return the IP addresses of attacker controlled nodes.
\item Compromising a DNS hosting account (e.g. by phishing) and changing DNS records to return the IP addresses of attacker controlled nodes.
\end{itemize}

\subsubsection{Data origin authentication of DNS seeds}

The aim of a DNS protocol level attack against a bitcoin DNS seed, such as a DNS cache poisoning attack, is to return a DNS response to the victim that points to nodes of the attackers choosing. The key security services that protect against this are:

\begin{enumerate}
\item{data integrity}: to detect whether data has been modified in transit\cite{everyday_crypto}.
\item{data origin authentication}: to confirm whether data came from a genuine sender\cite{everyday_crypto}.
\end{enumerate}

The principle method for applying data integrity and data origin authentication to DNS queries is the use of DNSSEC\cite{dnssec_rfc}. DNSSEC is a suite of specifications to extend the DNS protocol to allow DNS records and responses to be digitally signed by the owner of the domain.

At the time of writing (August 2017) there are currently six seed addresses listed in the bitcoin software\cite{bitcoin_chainparams}:

\begin{itemize}
\item seed.bitcoin.sipa.be
\item dnsseed.bluematt.me
\item dnsseed.bitcoin.dashjr.org
\item seed.bitcoinstats.com
\item seed.bitcoin.jonasschnelli.ch
\item seed.btc.petertodd.org
\end{itemize}

None of these addresses has DNSSEC configured and are therefore open to DNS protocol attacks such as DNS cache poisoning. 

\subsubsection{Control of DNS seed addresses}

It is not immediately obvious to users who controls the domain names associated with the DNS seeds. Also the mechanism that chooses the nodes that each seed will return is also not documented.

A brief analysis of public WHOIS records and bitcoin code repositories was carried out to determine the likely owners of the seed domains\cite{bitcoin_contributors}

Five out of the six domains are controlled by the primary bitcoin developers, who between them account for the vast majority of code contributions to the core bitcoin software\cite{bitcoin_contributors}. One is an academic author and researcher of bitcoin and has contributed to several papers referenced by this paper\cite{bitcoin_propagation}\cite{bitcoin_malleable}\cite{bitcoin_pay}.

Control of the DNS seeds appears to rest with six individuals, rather than a set of companies or institutions. This is perhaps a symptom of the decentralised, anarchic ideals evident in Nakamoto’s original paper\cite{bitcoin_original}, however whilst the source code of bitcoin is freely auditable, the operation of the DNS seed addresses and P2P network is not.

There are several security concerns, not specifically aimed at the current owners of the DNS seeds, but the principle of individuals controlling DNS seed domains. 

Firstly, that any individual may be influenced or may exploit their position of trust for personal gain. For example four of the six individuals are employed by Blockstream.com\cite{archive_luke}\cite{archive_decker}\cite{archive_wuille}\cite{archive_matt}, a private company developing blockchain based software and services.

Secondly there is a limit to the level of security any one individual can provide against a targeted attack from a highly resourced and motivated attacker. For example an attacker that compromised the DNS server or domain hosting account associated with a DNS seed could redirect all new nodes joining the network to their own malicious nodes.

Whether control of the bitcoin seed addresses places too much power in the hands of six bitcoin developers, is a question that poses both technical and philosophical considerations for bitcoin users. Much more broadly, it raises the question of who should have control of components of the bitcoin infrastructure that cannot be decentralised (such as DNS seeds) and how should those people or organisations be held accountable.

\subsubsection{Establishing connections}

Once a node (e.g. Node A) has learnt the IP address of another node (Node B) a connection is established by sending a VERSION message\cite{bitcoin_wiki_documentation} to Node B containing its software version number. 

If Node B is accepting connections from this particular software version it will reply with a VERACK message\cite{bitcoin_wiki_documentation}, which includes its own software version number.

Node A will send its own VERACK message if it also is accepting connections from this software version.

This exchange allows both nodes to check each other’s software versions before deciding to establish connections. Although it is not currently, this could provide a mechanism for node operators to exclude outdated software versions (with known security vulnerabilities) from participating in the network.

\subsubsection{Discovering nodes}

Once a node establishes a connection with another node (a peer), the node will query it for a list of network nodes that it is aware of by sending a GETADDR message\cite{bitcoin_wiki_documentation}. The peer will reply with an ADDR message\cite{bitcoin_wiki_documentation} containing a list of up to 1000 peers, randomly selected from the list of active peers that it is aware of.

Several academic studies\cite{bitcoin_topography}\cite{bitcoin_p2p}\cite{bitcoin_network}\cite{bitcoin_propagation} and projects\cite{bitcoin_nodes} have demonstrated that by sending GETADDR messages to each node, and subsequently sending GETADDR messages to every new node that is reported in ADDR messages, it is possible to discover all nodes currently active on the network.

This can be used to analyse the size and geographical distribution of the network \cite{bitcoin_network} as well as the environments that nodes are running in, such as cloud hosting providers, private datacentres or residential connections. 

In the context of security, as a node is identified by its IP address, this can in some circumstances be probabilistically linked to a physical location\cite{ip_geolocation}, \cite{ip_databases_unreliable}. On a broader scale, as nodes will freely report their software version in protocol handshakes (section 3.2.4), it is possible to scan the entire network for nodes running bitcoin software versions with known vulnerabilities. For example using the publicly available bitnodes.21.co project \cite{bitcoin_vulnerable}, at the time of writing (August 2017) there are 3 nodes running bitcoin core version 0.8.3, which is vulnerable to remote denial of service vulnerability CVE-2013-5700 \cite{bitcoin_exploit}.

\subsubsection{Transaction and block transmission}

The transmission of transaction and block information to all nodes is achieved with a broadcast mechanism. Once a node learns of a new transaction or block it is forwarded on to all its neighbours (the peers the node is actively connected to). These neighbours then forward the new transaction or block on to their neighbours and the process repeats until all reachable nodes in the network have received the new transaction or block.

Node A advertises the new transaction by sending an INV message \cite{bitcoin_wiki_documentation}, which includes a SHA256 hash of the new transaction (TXID) to Node B (see figure 3.4). 

If Node B is not aware of this new transaction ID (TXID) it will send a GETDATA message \cite{bitcoin_wiki_documentation}, which includes the TXID of the new transaction, to Node A. 

Node A will respond by sending a TX message \cite{bitcoin_wiki_documentation} containing the full transaction record to Node B.

\jfig{Advertising and transmitting a transaction between bitcoin nodes}{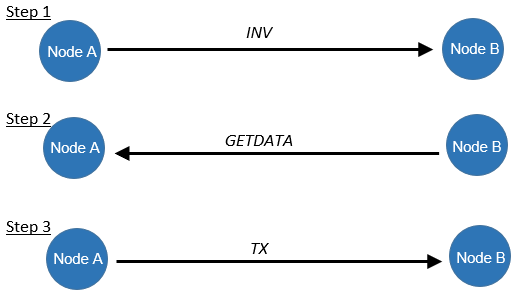}

Once Node B has successfully received the new transaction, it will validate it using its local copy of the blockchain and send INV messages to its neighbours to repeat the process.

The above process is the same for the transmission of blocks, except that the ID of the new block is sent in steps 1 and 2 (INV and GETDATA messages) and a BLOCK message \cite{bitcoin_wiki_documentation} is used to send the block information in step 3.

The example given assumes that only one transaction or block is being broadcast, when in reality INV and GETDATA messages can contain up to 50,000 transaction or block IDs.

\subsubsection{Requesting the latest blocks}

Whilst a node is online and connected to the network it will receive blocks as they are broadcast across the network and the node will keep its local copy of the blockchain up to date as new blocks arrive. However whilst a node is offline new blocks will have been created which the offline node will not be aware of, so upon reconnecting to the network the node will need to obtain the missing blocks. 

It will do this by sending a GETHEADERS message\cite{bitcoin_wiki_documentation} to a node that it is connected to, which includes the block ID of the last block that the node is aware of. 

For example (figure 2.3), Node A has just reconnected to Node B after being offline for some period of time. Node A will send a GETHEADERS message to Node B with a block ID of \jnumber{123}, which represents the last block in its local copy of the blockchain.

When Node B receives the GETHEADERS message it will compare the block ID \jnumber{123} with its local copy of the blockchain and if necessary reply with a HEADERS message\cite{bitcoin_wiki_documentation}, which contains the block ID’s of the remaining blocks in the chain. In this example it will include block ids \jnumber{124} and \jnumber{125} in its HEADERS message. 

Node A is now aware that there are two blocks, \jnumber{124} and \jnumber{125}, that it is missing.

\jfig{Querying a peer for the latest blocks}{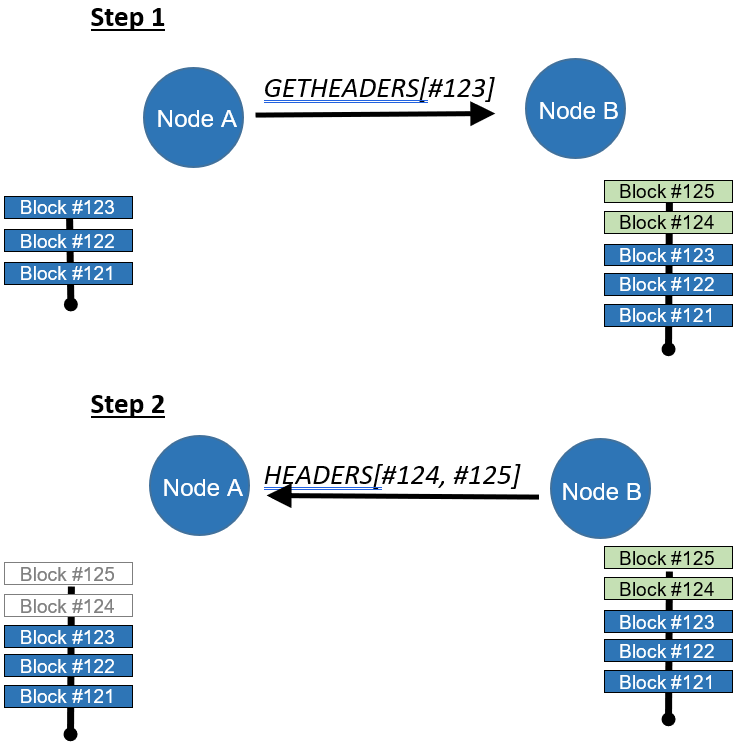}

Up to 2000 block IDs may be returned in a HEADERS message. If Node A needed to obtain more than 2000 blocks it would need to send out additional GETHEADERS messages.

The GETHEADERS/HEADERS exchange is designed to allow Node A to discover what blocks it is missing. The HEADERS message returned by Node B only contains block IDs, it does not contain the full block information. It is up to Node A to request each block with a series of GETDATA messages (see 2.2.6) to Node B, which will then send the full information for each block.

\subsubsection{Collecting unconfirmed transactions}

If a node is engaged in bitcoin mining activities it will need to collect and store unconfirmed transactions to include them in the block it is mining.

The collective pool of unconfirmed transactions within the network is called the ‘Mempool’ and miners will typically seek to gather as large a proportion of unconfirmed transactions as possible, in order to earn the most transaction fees once they discover a block. 

Nodes that have reconnected to the network after being offline have a method of gathering unconfirmed transactions from peers that they are connected to.

For example (figure 2.4), Node A has just reconnected to Node B after being offline for some period of time. Node A will send a MEMPOOL message\cite{bitcoin_wiki_documentation} to Node B to request a list of unconfirmed transactions that Node B is aware of. Node B will reply with an INV message containing the transaction IDs of all unconfirmed transactions that it is aware of.

\jfig{Retrieving a list of unconfirmed transactions}{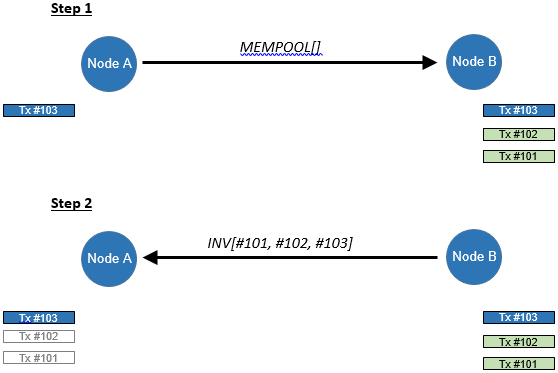}

Note that Node B does not know what unconfirmed transactions Node A is already aware of, so Node B will simply send an INV message containing all the unconfirmed transactions that it is aware of.

Once Node A has learnt of the outstanding transactions (\jnumber{101} and \jnumber{102}), it will then download the full transaction information from Node B using the GETDATA/TX message exchange.

\section{Network based vulnerabilities and attacks}

A distributed denial of service attack (DDoS) seeks to overwhelm a victim with more network traffic than the victim’s network connection or computing resources can cope with\cite{cert_smurf}.

One notable example of a DDoS attack is a DNS amplification attack\cite{cert_dns_amp}. Which in 2013 was used to generate 75Gbps of malicious traffic\cite{cloudflare_spamhaus} as part of a DDoS against Spamhaus\cite{spamhaus}, an organisation that combats email spam. At the time it was the largest DDoS attack ever seen.

The two properties that make this attack successful are

\begin{enumerate}
	\item Reflection: Responses are sent in reply to any incoming request, without authentication of the source\cite{ddos_reflect}.
	\item Amplification: The response is much larger than the initial request, meaning that the victim receives much more data than the attacker sends out\cite{ddos_reflect}.
\end{enumerate}

\subsection{Possible DDoS attacks}

Recall from section 2.2 that the bitcoin network protocol does not have any data origin authentication, but instead relies on the TCP sequence number to ensure that messages cannot be spoofed. However as mentioned there are numerous examples\cite{tcp_off}\cite{tcp_sequence}\cite{tcp_sequence2}\cite{tcp_offpath}\cite{challenge_response} where TCP sequence numbers do not provide adequate protection against spoofing.

Analysis was conducted on the bitcoin network protocol specification\cite{bitcoin_bitcoind} and the bitcoin core source code\cite{bitcoin_repo} for message exchanges that display the potential for reflection and amplification. Such properties would indicate the potential for their exploitation in DDoS attacks against bitcoin nodes.

\subsubsection{GETHEADERS/HEADERS}

Recall from section 2.2.7 that upon re-joining the network a node will seek to update its local copy of the blockchain by asking its peers for the blocks that have been created whilst it was away. It will send a GETHEADERS message with the block ID of the last block it is aware of and receive back a HEADERS message containing up to 2000 block IDs.

From examining the protocol specification\cite{bitcoin_wiki_documentation} and source code for processing an incoming GETHEADERS message\cite{bitcoin_getheaders} the standard response is to issue a HEADERS message to the listed source IP address. There are no data origin authentication or freshness checks included in the protocol specification or any method in the processing of a GETHEADERS message to determine whether the request is genuine.

Therefore it would imply that if an attacker was able to overcome the TCP sequence number, then a forged GETHEADERS message would result in a HEADERS message being sent to the victim. 

\jfig{GETHEADERS Reflection Attack}{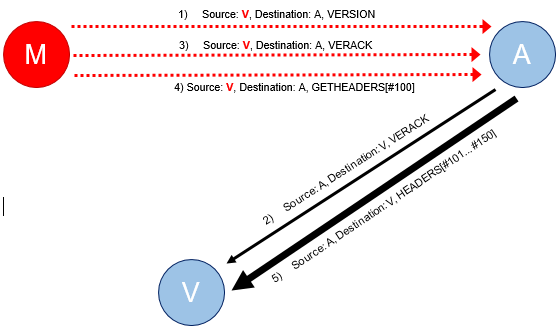}

For example (figure 3.1) M wants to induce A to send a HEADERS message to victim V. 

For A to accept the spoofed GETHEADERS packet, M must first trick A into establishing a connection with V.
M sends a VERSION message to A with the source address spoofed as V.
A sends a VERACK message to V.
M sends a VERACK message to A with the source address spoofed as V.

M has now established a connection with A on behalf of V and can now commence the DOS attack.
M sends a GETHEADERS message to A, with the source address spoofed as V, asking for blocks with an ID greater than \jnumber{100}.
A sends a HEADERS message to V with the block IDs for blocks \jnumber{101} to \jnumber{150}.

As well the potential for reflection, the GETHEADERS/HEADERS message exchange also has the potential for amplification.

The attacker must send the following messages:
1 x VERSION message with a size of 85 bytes\cite{bitcoin_wiki_documentation}.
1 x VERACK message with a size of 24 bytes\cite{bitcoin_wiki_documentation}.
1 x GETHEADERS message with a size of 69 bytes\cite{bitcoin_wiki_documentation}.

A total of 168 bytes.

Recall from section 3.2.9 that the resulting HEADERS message sent to the victim can contain up to 2000 block IDs, yielding a maximum message size of approximately 162,000 bytes or 158 Kilobytes. This is an increase by a factor of approximately 964.

\subsubsection{MEMPOOL}

Recall from section 2.2.8 that the any node can query another peer on the network for a list of all the unconfirmed transactions that it is aware of.

This is done by sending a peer a MEMPOOL message and in reply they will receive back an INV message containing a list of the transaction IDs (txID) of transactions that are unconfirmed. Recall from section 2.2.6 that an INV message can contain as many as 50,000 transaction IDs.

Again from examining the protocol specification \cite{bitcoin_wiki_documentation} and source code for processing an incoming MEMPOOL message\cite{bitcoin_mempool} the standard response is to issue an INV message to the listed source IP address. There are no data origin authentication or freshness checks included in the protocol specification or any method in the processing of a MEMPOOL message to determine whether the request is genuine.

Therefore it would imply that if an attacker was able to overcome the TCP sequence number, then a forged MEMPOOL message would result in an INV message being sent to the victim.

\jfig{MEMPOOL Reflection attack}{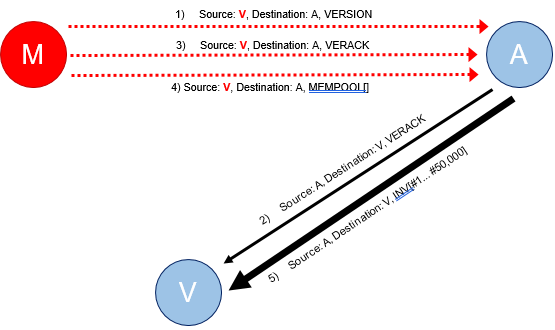}

For example (figure 3.2) M wants to induce A to send an INV message to victim V. 

In order for A to accept the spoofed MEMPOOL packet, M must first trick A into establishing a connection with V (section 3.2.4).

\begin{enumerate}
	\item M sends a VERSION message to A with the source address spoofed as V.
	\item A sends a VERACK message to V.
	\item M sends a VERACK message to A with the source address spoofed as V.\\
	M has now established a connection with A on behalf of V and can now commence the DOS attack.
	\item M sends a MEMPOOL message to A, with the source address spoofed as V, asking for a list of all unconfirmed transactions that A is aware of.
	\item A sends an INV message to V with the txIDs for transactions \jnumber{1} to \jnumber{50,000}.
	
\end{enumerate}

As well the potential for reflection, the MEMPOOL/INV message exchange also has the potential for amplification.

The attacker must send the following messages:

\begin{itemize}
	\item 1 x VERSION message with a size of 85 bytes\cite{bitcoin_wiki_documentation}.
	\item 1 x VERACK message with a size of 24 bytes\cite{bitcoin_wiki_documentation}.
	\item 1 x MEMPOOL message with a size of 24 bytes\cite{bitcoin_wiki_documentation}
\end{itemize}

A total of 133 bytes.

The size of the resulting INV message sent to the victim will vary, as although an INV message can contain up to 50,000 txIDs\cite{bitcoin_wiki_documentation} the actual number sent will depend on the number of unconfirmed transactions that node A is aware of at the time.

The number of unconfirmed transactions in the Mempool varies quite dramatically from hour to hour, for example on 28/6/17\cite{bitcoin_mempool_transactions} peaking at 28,758 and dropping to 6,607 within 4 hours. 

An attacker could track the size of the Mempool and time their attack to coincide with periods of high numbers of unconfirmed transactions, or indeed generate large amounts of unconfirmed transactions themselves. Therefore it is possible that an INV message may contain 50,000 transactions.

In evaluating the potential for amplification in this attack it is reasonable to assume that the resulting INV message sent to the victim will contain 50,000 txIDs. In which case, the size of the INV message will be approximately 1,800,000 bytes or 1.7 Megabytes. This is an increase by a factor of approximately 13,534.

The precise amount of data that the attacker sends and the victim receives will vary, as each message will be encapsulated in an ethernet frame (20 bytes\cite{ethernet}), an IP packet (36 bytes\cite{ip}) and a TCP segment (20 bytes\cite{tcp}), so a total of 145 bytes is added to the size of each message.

However the potential for reflection and the scale of the amplification leads to the conclusion that the GETHEADERS and MEMPOOL message exchanges have the potential for exploitation in a denial of service attack. This hypothesis should be confirmed through experimentation
\subsection{Comparison to other amplification vectors}

While a factor of 13,534 is quite powerful, there are many other methods that can be used to amplify the size of a DDOS attack. One example is using Memcached, which has achieved a factor of 51,200\cite{cloudflare_memcached} in practical attacks. This also has the additional benefit that it can be created on demand, as an attacker can place large objects on the server without paying transaction fees, rather than the waiting an attacker would need to perform the same attack using bitcoin if they didn't want to pay. Memcached was used in the  attack against OVH\cite{ovh_attack} which managed to hit 1.7tbps, which was helped by the number of open servers which were used.

The fact that there are other amplification vectors available does not mean that the Bitcoin ones should not be fixed though, firstly because when the other vectors are fixed the bitcoin ones will become a reasonable vector to use. There is also the fact that memcached traffic crossing a large network should be quite rare, as the overhead would be very high, making detection easier, whereas the same attack with bitcoin would raise less suspicion from the network of a single amplifier.
\section{Possible Solutions}

As explained in section 2.2.4, it is trivial to discover the versions of node software used across the network and to identify nodes running software with known vulnerabilities. 

Currently the node software will still establish connections with nodes running outdated versions, however the connection handshake could provide a mechanism to exclude these nodes from the network until they are updated. A minimum version number could be used as a criteria for accepting a connection thereby enforcing a minimum version necessary to participate in the network.

Recall from section 3.2 that there is the potential for two message exchanges (GETHEADERS/HEADERS and MEMPOOL/INV) to be used to induce a target node to send data to a victim node, potentially leading to a DDoS attack. 

This is because once the TCP sequence number is defeated, the bitcoin connection handshake (figure 3.3) can be spoofed which would in turn allow the GETHEADERS or MEMPOOL messages to be spoofed. Preventing spoofing of the handshake would provide protection against subsequent message exchanges from being abused.

One method of preventing an off-path attacker from spoofing the connection handshake would be to add a random nonce to the VERACK message.

\jfig{Bitcoin connection handshake with a random nonce added to the VERACK messages}{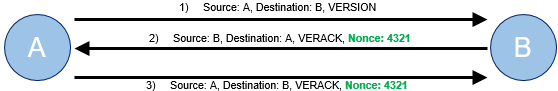}

For example (figure 4.1), Node A attempts to establish a connection with Node B:

\begin{enumerate}
\item Node A sends a VERSION message to Node B.
\item Node B responds to Node A with a VERACK message, which includes a randomly generated nonce.
\item Node A completes the handshake by sending its own VERACK message back to Node A, including the nonce it received from Node B. 
\item Node B checks that the nonce received in step 3 is correct and if so it will accept and process future messages from Node A. Otherwise Node B will ignore messages from Node A until a correct handshake is completed.
\end{enumerate}

\jfig{An attempted spoof connection handshake with a random nonce}{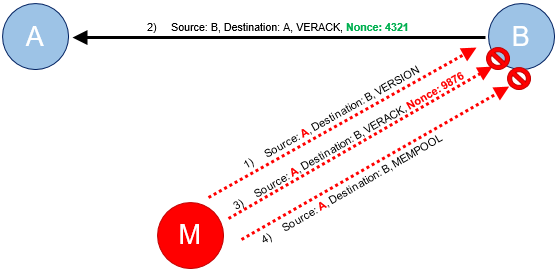}

In figure 4.2 malicious Mode M tries to spoof a connection attempt from Node A with Node B.

\begin{enumerate}
\item Node M sends a VERSION message to Node B, changing the source address to appear that it came from Node A.
\item Node B responds to Node A with a VERACK message, which includes a randomly generated nonce. As this VERACK message is routed to Node A, Node M will not learn the value of the nonce.
\item Node M attempts to complete the handshake by sending a VERACK message  to Node B, but it is forced to guess the value of the nonce generated in Step 2. Assuming that the nonce value is chosen in a secure way (explained below) it is very unlikely that the correct nonce value will be guessed.
\item Any subsequent messages received by Node B with a source address of Node A (such as a spoofed MEMPOOL message) will be rejected until a correct handshake is completed.
\end{enumerate}

To ensure that a nonce value cannot feasibly be guessed, it should:

\begin{itemize}
\item Be randomly generated using a suitable pseudorandom number generator.
\item Be sufficiently large to make brute force guessing infeasible, for example at least 32-bits (the size of the TCP sequence number\cite{tcp}).
\end{itemize}

In order to detect attempted abuses of the bitcoin connection handshake the node software should also log and report incomplete connection handshakes. For example if incorrect nonces are repeatedly being given by a node this might indicate a spoof connection attempt.

\section{Conclusion}

To operate as a currency Bitcoin requires a P2P network to function. The operation of this P2P network is not well documented or widely studied and yet plays a crucial role in maintaining the overall security of the currency.

Users of Bitcoin rely solely on cryptography to establish trust, not on the authority of a trusted third party but on mechanisms that users can use to validate transactions themselves. Therefore this principle should also apply to communications on the P2P network – the actions and messages of other nodes should not be trusted and should be assumed to be malicious. Translating this principle into the design of bitcoin’s network protocols and software does not appear to have happened.

In conclusion, the key contribution of this project are the following recommendations:

\begin{enumerate}
	\item Protocol hardening: The bitcoin network protocol and core software implementation requires a thorough security audit, to address potential security vulnerabilities identified in the HEADERS and MEMPOOL message exchanges. The use of network security mechanisms such as random nonces, cryptographic integrity protection and entity authentication should be further considered. For example the introduction of a random nonce during the connection handshake could provide protection against message spoofing.
	
	\item Management of network infrastructure: Parts of bitcoin’s network architecture are not decentralised (e.g. DNS seed addresses) and remain under the control of individual users (section 2.2.3). This holds the potential for conflicts of interest or insider attacks and bitcoin users need to consider how components of the bitcoin infrastructure that cannot be decentralised are governed and managed in an accountable and transparent way.
	
\end{enumerate}

\subsection{Further work}

Several areas where further work could be carried out have been identified.

Firstly the potential for bitcoin message spoofing should be investigated through experimental analysis. In particular the potential for the use of HEADERS and MEMPOOL messages in DDoS attacks should be evaluated in a lab environment as well as the live bitcoin network.

Secondly, mechanisms for providing data integrity and entity authentication for all network nodes should be incorporated into the network protocol. The current proposals\cite{bitcoin_bip} only provide these services for a sub set of nodes that coordinate between themselves and cannot be used at scale across the whole network.

There is also a need to harden Bitcoin against future attacks, such as quantum ones, which may become a problem around 2027\cite{bitcoin_quantum}. This is the point at which quantum computes will be able to be fast enough to challenge the current generation of ASIC based miners.
\bibliography{main}{}
\bibliographystyle{unsrt}

\end{document}